\documentclass{aa}
\usepackage{natbib}
\usepackage{amsmath}
\usepackage{amssymb}
\usepackage{graphicx}
\usepackage{txfonts}
\usepackage{float}
\bibpunct{(}{)}{;}{a}{}{,}          

\begin{document}
\title{Temporal comparison of nonthermal flare emission and magnetic-flux change rates}

   \author{C.~H.~Miklenic\inst{1,2} \and
              A.~M.~Veronig\inst{2} \and
              B.~Vr\v{s}nak\inst{3}
          }

   \offprints{C.~H.~Miklenic, \email{christiane.miklenic@uni-graz.at}}

   \institute{Space Research Institute, Austrian Academy of Sciences,
   Schmiedlstra\ss e 6, A--8042 Graz, Austria \and
   Institute of Physics, University of Graz, Universit\"atsplatz 5,
    A--8010 Graz, Austria \and
Hvar Observatory, Faculty of Geodesy, Ka\v ci\'ceva 26,
        HR--1000 Zagreb, Croatia
             }

   \date{Received 10 September 2008 / Accepted 2 February 2009}


\abstract
     {To understand the mechanisms that trigger solar flares, we require models describing and quantifying observable
responses to the original energy release process, since the coronal energy release site itself cannot be
resolved with current technical equipment. Testing the usefulness of a particular model requires the comparison of its predictions with flare observations.}
     {To test the standard flare model (CSHKP-model), we measured the magnetic-flux change rate
     in five flare events of different GOES classes using chromospheric/photospheric observations and compared its progression with observed nonthermal flare emission. We calculated the cumulated positive
     and negative magnetic flux participating in the reconnection process, as well as
     the total reconnection flux. Finally, we investigated the relations between the total reconnection flux, the GOES class of the
 events, and the linear velocity of the flare-associated CMEs.}
     {Using high-cadence H$\alpha$ and TRACE~1600~{\AA} image time-series data and MDI/SOHO magnetograms, we
     measured the required observables (newly brightened flare
     area and magnetic-field strength inside this area). RHESSI and INTEGRAL hard X-ray time profiles in nonthermal energy bands
     were used as observable proxies for the flare-energy release rate.}
     {We detected strong temporal correlations between the derived magnetic-flux change rate and
     the observed nonthermal emission of all events. The cumulated positive and negative
     fluxes, with flux ratios of between 0.64 and 1.35, were almost equivalent to each other.
     Total reconnection fluxes ranged between $1.8 \times 10^{21}$~Mx for the weakest
     event (GOES class B9.5) and $15.5\times 10^{21}$~Mx for the most energetic one
     (GOES class X17.2). The amount of magnetic flux participating in the reconnection process was higher in more energetic
     events than in weaker ones. Flares with more reconnection flux were associated with faster CMEs.}
     {}


\authorrunning{C.~H.~Miklenic et.al.}
\titlerunning{Nonthermal flare emission and magnetic-flux change rates}
\maketitle

\section{Introduction}

Solar flares are intriguing, intensely-studied phenomena. Much progress has been made in understanding the processes that occur on the Sun when a flare occurs. However, these events remain a partly-unsolved mystery. It is generally accepted that the energy fueling a solar flare is magnetic energy, which is accumulated slowly and stored inside the corona. After its release by magnetic field reconnection, it is converted into the kinetic energy of fast particles, plasma flows, heat, and MHD shock waves. Unfortunately, our observations remain insensitive to the primary energy-release process, since the energy release site in the reconnecting current sheet cannot be resolved with our
current technical equipment. To learn about the properties of the energy release site and the processes occurring there, we therefore depend on exploring the plasma responses to the original energy-release process.

One of the observable consequences of the energy release in the corona are the bright, separating flare ribbons, which can be seen in UV bands and chromospheric spectral lines, and are prominent in H$\alpha$. They are located on each side of
the magnetic polarity inversion line and move away from it during the course of the flare. This apparent motion can be explained by the most widely accepted model of erupting flares, the CSHKP model \citep{car64,sturr66,hir74,kopp76}. According to this, the separating flare ribbons are the chromospheric signatures of the energy release in the corona. The energy, which has been released at the primary energy-release site, is transported downward into the chromosphere along the newly reconnected field lines, by fast particles and thermal conduction. When it is deposited at the two footpoints of the field lines, which are rooted inside opposite magnetic polarities, the chromosphere brightens, i.e., the so-called flare kernels are created, and many of these adjacent kernels form elongated flare ribbons. The reconnection site is located below the flare-associated erupting filament/CME, and, since the erupting structure rises continuously to higher coronal heights, the reconnection also occurs at successively higher altitudes, as the flare progresses. As a consequence, field lines with footpoints rooted further and further away from each other are involved in the reconnection process. Thus, the newly brightened areas of both ribbons are located further and further apart and we observe separating flare ribbons.

Since the coronal reconnection site is coupled with the chromosphere by the reconnected field lines, it is obvious that the coronal magnetic reconnection rate, i.e., the rate at which magnetic field lines are reconnected, is associated
with the separating flare ribbons. Based on the CSHKP model, \citet{forbes84} and \citet{forbes00} considered the rate of photospheric magnetic flux-change $\dot{\varphi}$. Using magnetic flux conservation, they derived a quantitative
estimate for the coronal reconnection rate or magnetic flux change rate $\dot{\varphi}$, respectively, that can be derived from observable quantities

\begin{equation}
\dot{\varphi}=\frac{\partial}{\partial t} \int B_n\,da,
\end{equation}
where $B_n$ is the photospheric magnetic-field strength component perpendicular to the solar surface inside the newly brightened area $da$, which is swept up by the flare ribbons. We denote by $\dot{\varphi}$ the rate at which magnetic flux from formerly separated magnetic domains is unified. The original CSHKP model is a 2D model, since it describes the evolution of a flare in a vertical plane. In the third dimension, namely, the direction of the polarity inversion line, translational symmetry is assumed. This assumption converts the model to a 2$\frac{1}{2}$D model. It is likely, however, that the extension in the third dimension is not continuous, but rather highly fragmented into temporary
magnetic islands. However, we note that calculating the magnetic flux change rate $\dot{\varphi}$ does not require the assumption of translational symmetry. Therefore, $\dot{\varphi}$ is valid in three dimensions.

To test the CSHKP model, an observable quantity is required that can be regarded as a proxy for the energy release rate or reconnection rate, respectively, which we are incapable of measuring directly. The fast electrons that spiral downward
along the newly reconnected field lines, generate microwave gyrosynchrotron emission, and when they deposit their energy at the footpoints of the reconnected field lines, hard X-ray (HXR) emission is produced by the nonthermal bremsstrahlung of electrons being scattered off ions. Therefore, the observed microwave and HXR fluxes act as indicators of the rate of
accelerated electrons, which carry a large fraction of the total energy released during a flare \citep[e.g.,][]{hud91,dennis03}. Thus, microwave and HXR emission can be used as proxies for the flare energy-release rate. If the CSHKP model is applicable, the derived magnetic-flux change rate $\dot{\varphi}$ should exhibit a similar temporal evolution as the observed
HXR and microwave flux, i.e., the maxima in the $\dot{\varphi}$-profile should coincide approximately co-temporally with peaks in the observed nonthermal flare-emission. Strong temporal correlations of observed nonthermal emission with derived quantities characterizing the magnetic reconnection process, such as the magnetic flux change rate, the electric field at the reconnecting X-point, or the Poynting flux, which is transported into the
reconnection region, were found in several cases \citep[e.g.,][]{asa04,qiu04,jing05,isobe05,lee06,temmer07,mikl07}.

It is known from observations that the energy release in a dynamical flare is also related to the kinematics of the associated CME \citep[e.g.,][]{maricic07,temmer08}. Statistical studies indicate that CME parameters, such as the velocity and kinetic energy, are correlated with the soft X-ray (SXR) peak flux and the time-integrated SXR flux, i.e., the flare fluence \citep[e.g.,][]{moon02,bojan05c}. \citet{qiu05} reported that a greater amount of total reconnection flux is related to higher CME velocities.

In this paper, we present a detailed analysis of five well-observed erupting flare events of
different GOES classes. We calculated $\dot{\varphi}$ to test the model prediction, namely, the co-temporal evolution of the derived magnetic-flux change rate and observed nonthermal flare emission. In addition, we determined the
cumulated reconnection-flux profiles for the positive and negative magnetic polarity domain as a function of time, the ratio of these fluxes, as well as the total flux that had to have been reconnected until the end of the flare.
Finally, we investigated the relations between the total reconnection flux, the
GOES class of the events, and the linear velocity of the flare-associated CMEs.

The paper is structured as follows. Section~\ref{sec:dobs} contains descriptions
of the used data sets, Sect.~\ref{sec:analysis} describes the methods that we have applied, Sect.~\ref{sec:results} lists the results for each flare, and in
Sect.~\ref{sec:discussion} the results are summarized and discussed.

\section{Data and observations}\label{sec:dobs}

\begin{figure*}
\resizebox{\hsize}{!}{\includegraphics{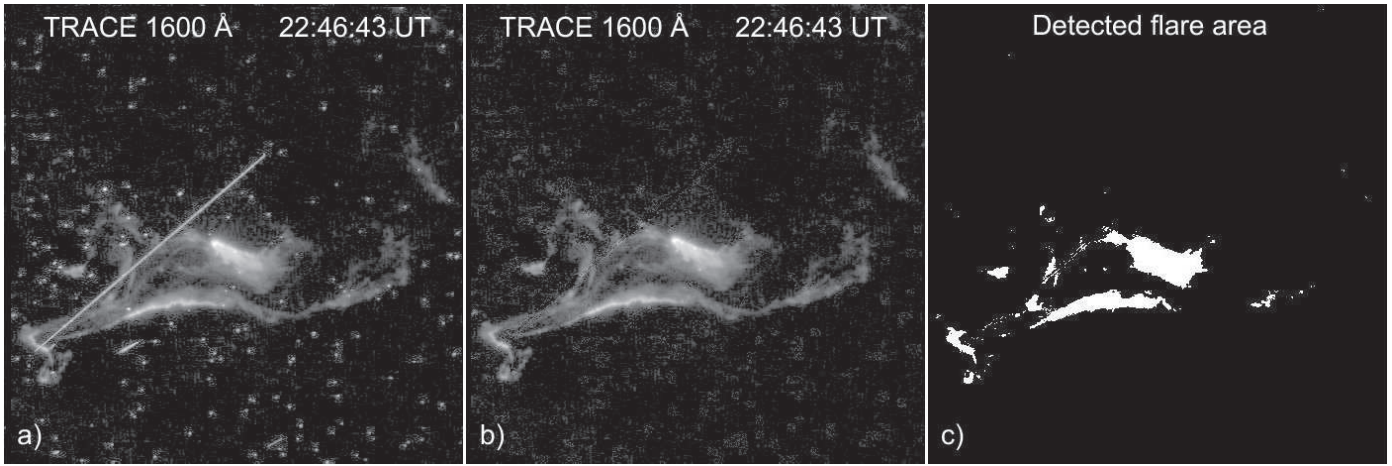}} \caption{2005 January 15,
X2.6 flare~--~\textbf{a)}: Image frame corrupted by many particle hits. \textbf{b)}: The same image after the particle hits have been removed by the running median method. \textbf{c)}: Flare area at 22:46:43~UT detected from b).}
\label{fig:07b}
\end{figure*}

\begin{table*}
\begin{center}
\caption{Event Information. For each event we list the date, NOAA active region, position, and GOES class plus GOES maximum of the flare, as well as the first LASCO C2 appearance, the central position angle, and the linear velocity of the associated CME. Details on the CMEs were obtained from the SOHO LASCO CME CATALOG \citep[][~--~\texttt{http://cdaw.gsfc.nasa.gov/CME\_list/}]{yashiro04}.}
\label{tab:1}
\begin{tabular}{c | c l c c | c c c}
  \hline \hline
  & \multicolumn{4}{c}{Flare} & \multicolumn{3}{|c}{CME}\\ 
 Date   & NOAA  & Position & \multicolumn{2}{c|}{GOES} & first LASCO C2 & central PA & linear velocity\\
   & AR
   &  & Class & Max. [UT] &  appearance [UT] & [deg] & [$\rm{km\,s^{-1}}$]\\
  \hline
  2003 Oct 28 & 10486 & S16 E08 & X17.2 & 11:10 & 10:54  & Halo & 2460\\
  2003 Nov 18 & 10501 & S02 E37 & M3.9 & 08:31  & 08:50 & Halo  & 1660\\
  2005 Jan 15 & 10720 & N13 W04 & X2.6 & 23:02  & 23:06 & Halo & 2860\\
  2006 Jul 06 & 10898 & S10 W30 & M2.5 & 08:35  & 08:54 & Halo &  910\\
  2007 May 19 & 10956 & N01 W05 & B9.5 & 13:02  & 13:24 & 260 & 960\\
  \hline
\end{tabular}
\end{center}
\end{table*}

\begin{table*}
\caption{Information on the used data sets. For each event we list the instrument, wavelength, cadence, and pixel scale of the image time series. A 'yes' in the Saturation-column indicates that the observations were saturated during the impulsive phase of the event. The last two columns give the used instrument and energy band of the observed nonthermal HXR emission.} \label{tab:2}
\begin{tabular}{c|c c c c c |c c}
  \hline \hline
   &   \multicolumn{5}{|c|}{Image time series} &  \multicolumn{2}{|c}{HXRs} \\
  \raisebox{1.2ex}[-1.2ex]{Date} &  Instrument  & Wavelength &
    Cadence [s] & \arcsec/pixel &
     Saturation &  Instrument & Energy [keV] \\
\hline
   2003 Oct 28 & Meudon Obs. & H$\alpha$ blue wing &  $\sim 60$  & 1.9  & no  &
     INTEGRAL/SPI & $>$ 150\\
  2003 Nov 18 & TRACE & 1600~{\AA} & $\sim 23$ & 0.5 & no & RHESSI & 20~--~60\\
  2005 Jan 15 & TRACE & 1600~{\AA} & $\sim 10$ & 0.5 & no &  RHESSI & 30~--100\\
   & BBSO & H$\alpha$ line center & $\sim 60$ & 1.05 & yes &  & and 100~--300\\
  2006 Jul 06 & Hvar Obs. & H$\alpha$ line center & $\sim 4$ & 0.3 & yes & 
    RHESSI & 20~--~50\\
  2007 May 19 & KSO & H$\alpha$ line center & $\sim 60$ & 2.2 & yes &  RHESSI & 15~--~50\\
  \hline
\end{tabular}
\end{table*}

The five flare events presented in this paper were eruptive (two-ribbon) flares,
associated with a CME. Table~\ref{tab:1} overviews the flare/CME events. In each
flare, two ribbons are clearly shaped, and in some cases also a third ribbon or distant
brightenings can be seen. The position of a flare on the solar disk served as a selection criterion, since the magnetic field component normal to the solar surface is required for the analysis (see Sect.~\ref{sec:analysis}). Therefore, we chose only events that were located not more than $\pm 40^\circ$ away from the disk center. A second selection criterion was the availability of high-cadence image time-series data in H$\alpha$ or UV and complete coverage of the impulsive phase in HXRs.

To analyze the flare events, we used the following data sets (see also summary in Table~\ref{tab:2}):

\begin{enumerate}
\item Full-disk line-of-sight magnetograms, provided by the SOI/MDI
instrument \citep{scherr95} onboard the Solar and Heliospheric Observatory
(SOHO). In all events except for that on 2003 November 18, five successive magnetograms one-minute apart were averaged
together to form a single low-noise magnetogram, which was used in further analysis.

\item High-cadence (4~--~60~s) image time-series data in H$\alpha$ or UV. The H$\alpha$ images
were provided by various ground-based observatories: Kanzelh\"ohe Solar
Observatory \citep[KSO, Austria;][]{otruba03}, Big Bear Solar Observatory
\citep[BBSO, USA;][]{denker99}, Hvar Observatory \citep[Croatia;][]{otruba05}, and
Meudon Observatory (France). UV images in the 1600~{\AA} passband were obtained from the Transition Region and Coronal Explorer \citep[TRACE;][]{handy99}.

\item Full-disk, nonthermal HXR-intensity time-profiles from the Reuven Ramaty High
Energy Solar Spectroscopic Imager \citep[RHESSI;][]{lin02}. For one
event (2003 Oct 28), HXR observations from the gamma-ray spectrometer SPI onboard the International Gamma-Ray
Astrophysics Laboratory \citep[INTEGRAL;][]{integral} were used. The energy bands
of the RHESSI profiles were chosen in such a way that the emission was clearly
nonthermal (steep rise~--~fast decay of the peaks). At the same time, the photon
energies had to be low enough to produce reasonable count statistics. Since the
GOES classes of the analyzed events ranged between B and X, different energy bands
of the RHESSI profiles were used in each case to satisfy these demands.
\end{enumerate}

\emph{Data reduction}: All images of a particular event were rotated to solar
north, if necessary, and differentially rotated to the same reference time.
H$\alpha$ images were cross-correlated in time to account for seeing effects.
Coalignment of the different data sets was accomplished using MDI continuum,
H$\alpha$ blue wing, and TRACE~WL images. Sunspots near the flare sites were taken
as a reference for coalignment by cross-correlation techniques. The different
pointings of the TRACE~WL and 1600~{\AA} telescopes were taken into account. MDI
images were converted from SOHO-view to Earth-view.

When using intensity thresholds to differentiate between flare pixels and
non-flare pixels, transient bright non-flare features, such as cosmic rays, will inevitably be included among
ribbon pixels. This effect was distinct during the impulsive phase of one of the events studied (2005 Jan 15, X2.6). Since these features do not survive for more than 1 or 2 frames at any given pixel, they can easily be eliminated. One possibility is to create running mean images over a few frames. Thus, bright particle hits are smoothed out \citep[e.g.,][]{longcope07,qiu07}, although, the intensity of the flare pixels is also modified. Therefore, we created running median images in the image time series of this event. For each pixel $x$, we calculated the median of frames $i$ to $i+n$ (with $n=5$) at the location of $x$. If $x$ is a
flare pixel, it will also be bright in the following frames and the median is
high. If pixel $x$ is a particle hit, however, it will be dark again in the subsequent
frames, and the median will be low. If the median is lower than a particular intensity
threshold, pixel $x$ in frame $i$ will be replaced by the median, whereas all pixels
with medians exceeding the threshold remained unchanged. Thus, only particle hits
were removed from the images, while the intensity of the flare pixels was not
modified (cf.~Fig.~\ref{fig:07b}). To check this method, we calculated the total
flare area, derived from both the original time series of this event and the frames, where the
particle hits had been removed, and found a total flare-area overestimation of
$\sim$100\%, when using the original times series. The overestimation of the
reconnection flux, however, was only $\sim$20\%, since particle hits cover the
entire FOV, i.e., also regions far away from the flare site, where the magnetic
fields are weak.

\section{Analysis}\label{sec:analysis}

To determine the magnetic-flux change rate $\dot{\varphi} =
\frac{\partial}{\partial t}\int B_n\,da$ we measured the following quantities:

\begin{enumerate}
\item The newly brightened area $da$ in an image compared with the preceding images, separately for each magnetic polarity domain. This was done by: (1) creating base difference images from the original time series; (2) determining the
smallest intensity maximum $I_{sm}$ of the entire difference-image time-series; (3) multiplying $I_{sm}$ by various scaling factors to derive a set of potential intensity thresholds. Thresholds that enabled us to differentiate correctly between flare pixels and non-flare pixels, were considered to be appropriate. The usability of the various threshold candidates was estimated by considering the total flare area detected in using them. Thresholds that were too low produced total flare areas
that also contained non-flare pixels, whereas relatively high thresholds omitted fainter parts of the flare ribbons. Since several thresholds within a certain range seemed to be appropriate for identifying the total flare area, the analysis
was carried out for a range of reasonable thresholds, and then the mean of the thus calculated magnetic-flux change-rate profiles was taken. We found that the $\dot{\varphi}$-peak-times were insensitive to the different thresholds, but the height of the peaks changed by 5~--~15\%.

To be counted as a member of $da$, a particular pixel had to exceed the given threshold, and be a non-flare pixel in the preceding images. Furthermore, it had to be located inside the currently-analyzed magnetic-polarity
domain, and exceed the MDI noise level of 20~G.

\item The normal component of the photospheric magnetic-field strength $B_n$ inside $da$. The line-of-sight magnetograms of MDI are known to be less sensitive to fields stronger than approximately 1700~G, i.e., MDI underestimates
strong fields \citep{berger03}. Following the cross-calibration study of \citet{berger03}, we multiplied the reported MDI line-of-sight magnetic field values by 1.56. Furthermore, we assumed that the field is
approximately radial at the photosphere, and therefore divided the photospheric magnetic field by the cosine of the central meridional distance of the flare to derive the radial magnetic-field strength at each pixel.
\end{enumerate}

At each time $t$, we measured both the newly brightened area $da$, consisting of all pixels meeting the aforementioned criteria, and $B_n$ at the locations of these pixels, and then calculated the positive and negative magnetic
reconnection flux at each time by multiplying the pixel area and $B_n$ of each pixel and adding the products. Division of the reconnection flux by the time intervals between two consecutively taken images yielded the magnetic flux change-rate for the positive and negative polarity domain, $\dot{\varphi}_+(t)$ and $\dot{\varphi}_-(t)$, respectively. The magnetic flux change-rate
$\dot{\varphi}$ was calculated by taking the mean of $\dot{\varphi}_+$ and $\dot{\varphi}_-$.

We also determined the cumulated magnetic reconnection flux at each time for both polarity domains, $\varphi_+(t)$ and $\varphi_-(t)$, respectively, by adding the newly reconnected flux at time $t$ to the flux that had been reconnected up to
time $t$. Since equal amounts of positive and negative magnetic flux are involved in the reconnection process at each time, the $\varphi_+(t)$ and $\varphi_-(t)$-profiles should be identical in the ideal case. We also calculated
the total reconnection-flux profile $\varphi_{\rm{tot}}(t)$ to be the mean of $\varphi_+$ and $\varphi_-$, and we used the $\varphi_{\rm{tot}}$-profiles obtained from the lowest and highest of the appropriate thresholds as an error estimate of the total
reconnection flux.

\section{Results}\label{sec:results}

The results are presented separately for each event. The following is valid
for all events. Panels (a)~--~(d) of Figs.~\ref{fig:01}, \ref{fig:02},
\ref{fig:03}, \ref{fig:04}, and \ref{fig:05} show the evolution of the flare
ribbons in each event. Panels (a) were taken during the impulsive phase, panels
(b) show the situation at the time of the maximum of the HXR profiles, panels (c)
display the flare ribbons at the time of GOES maximum, and panels (d) are decay-phase images. Panels (e) show the calculated total flare area, i.e., the sum of the newly brightened areas in all images, superposed on the decay-phase image, and in panels (f), the contours of this area are plotted on the MDI magnetogram of the flaring region. The top panels of Figs.~\ref{fig:01a}, \ref{fig:02a}, \ref{fig:03a}, \ref{fig:04a}, and \ref{fig:05a} show the GOES12~1~--~8~{\AA} soft X-ray (SXR) flux, as well as the  cumulated reconnection-flux profiles, and in the bottom panels, the derived magnetic-flux change rate $\dot{\varphi}$ is presented along with the observed nonthermal HXR emission profile.

\subsection{2003 October 28, X17.2 flare}\label{subsec:28-Oct-2003}

\begin{figure}[h]
\resizebox{\hsize}{!}{\includegraphics{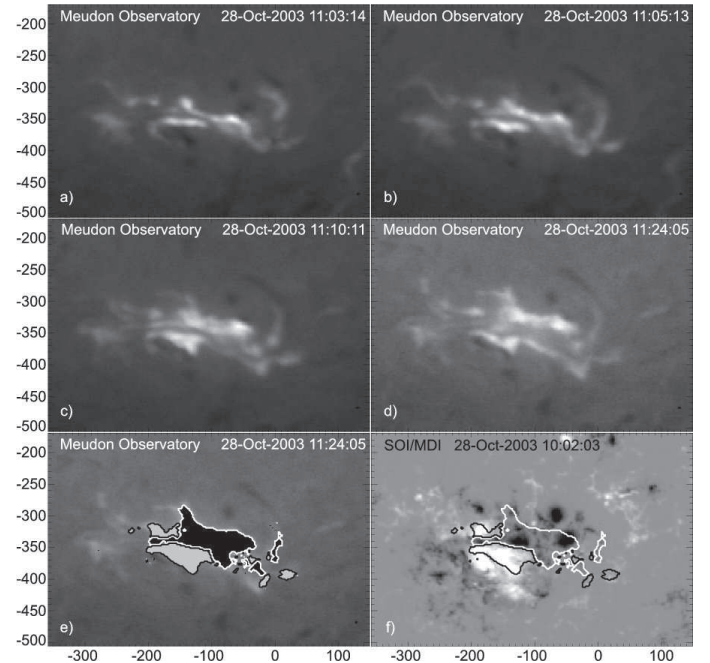}} \caption{2003 October 28, X17.2 flare~--~\textbf{a)~--~d)}: Temporal evolution of the flare ribbons in
H$\alpha$. \textbf{a)}: impulsive phase image, \textbf{b)}: time of INTEGRAL/SPI
maximum, \textbf{c)}: time of GOES maximum, \textbf{d)}: decay phase image.
\textbf{e)}: calculated total flare area superposed on decay phase image (black with
white contours: negative polarity, gray with black contours: positive polarity),
\textbf{f)}: total flare area on MDI magnetogram. Contours are the same as in e).
MDI data range scaled to $\pm 1000$~G out of [$-1680, +1650$]~G.~~--~FOV: $505\arcsec
\times 334\arcsec$.} \label{fig:01}
\end{figure}

Panels (a)~--~(d) of Fig.~\ref{fig:01} show the flare evolution in Meudon
H$\alpha$, and panel (e) displays the total flare area. It consists of the two
large regions, swept up by the separating flare ribbons, as well as some
smaller regions in the vicinity that also brightened up in the course of the
flare. In Fig.~\ref{fig:01}~(panel~f), the contours of the total flare area are plotted on
the MDI magnetogram. The northern ribbon passed through the negative magnetic
polarity domain, whereas the southern one appeared in the positive polarity. Parts
of both ribbons passed regions of magnetic field strength exceeding 1500~G.

\begin{figure}[h]
\resizebox{\hsize}{!}{\includegraphics{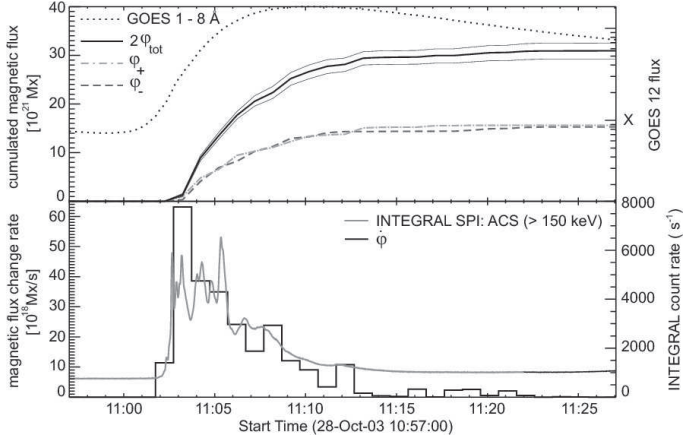}} \caption{2003 October 28, X17.2 flare~--~\textit{Top:} GOES12~1~--~8~{\AA} soft X-ray (SXR) profile,
cumulated total magnetic reconnection flux ($\varphi_{\rm{tot}}$) plus error estimate,
and cumulated positive and negative reconnection flux ($\varphi_+$, $\varphi_-$).
\textit{Bottom:} INTEGRAL SPI HXR countrate and magnetic flux change rate
($\dot{\varphi}$).} \label{fig:01a}
\end{figure}

The top panel of Fig.~\ref{fig:01a} shows the GOES flux, as well as the cumulated total
reconnection flux along with its error estimate. For the sake of clarity, we plot
$2\,\varphi_{\rm{tot}} = \varphi_+ +\varphi_-$. In addition, the cumulated
reconnection flux originating separately from the positive and negative magnetic polarity
domains, $\varphi_+$ and $\varphi_-$, is presented. The positive and negative
flux profiles look similar over the entire time range. This indicates that
almost equal amounts of positive and negative magnetic flux participate in the
reconnection process at a given time, as theoretically expected. During the
impulsive phase, when more and more flux is reconnected, the cumulated flux
profiles steeply rise. In the decay phase, i.e., after the GOES flux reached its
maximum, the reconnection process slowly comes to an end, and the amount of newly
reconnected flux decreases. This results in nearly constant cumulated flux
profiles during this phase of the event. At the end of the analyzed time interval,
the ratio of cumulated positive versus negative reconnection flux is 1.02, and the
total flux $\varphi_{\rm{tot}}$ adds up to $\sim 15.5 \times 10^{21}$~Mx
(cf.~Table~\ref{tab:3}).

The INTEGRAL SPI count-rate of $> 150$~keV, and the derived magnetic flux
change-rate are presented in the bottom panel of Fig.~\ref{fig:01a}. The INTEGRAL
profile exhibits several spikes embedded in a broad peak, which lasts from
$\sim$~11:02~UT to 11:10~UT. Since the cadence of the H$\alpha$ images was only
about 1~min, it was not possible to resolve every single INTEGRAL spike, although, the magnetic flux change-rate does reflect
the overall shape of the INTEGRAL flux.

\subsection{2003 November 18, M3.9 flare}\label{subsec:18-Nov-2003}

This event was already published in \citet{mikl07}. We present it again here, because the cumulated magnetic reconnection-flux profiles had not then been calculated, but only the magnetic flux ratio of the overall positive to negative reconnection flux, taken at the end of the analyzed time interval.

\begin{figure}[h]
\resizebox{\hsize}{!}{\includegraphics{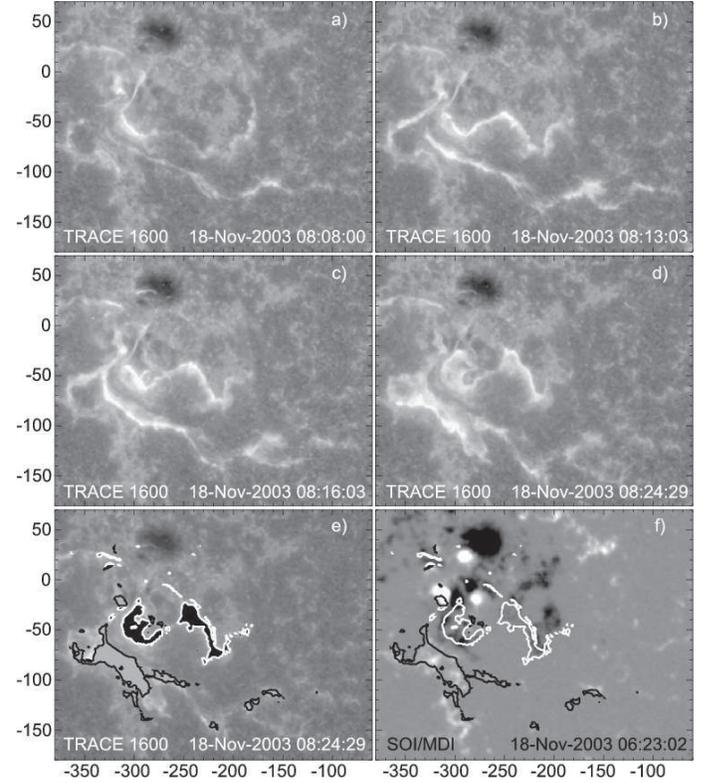}} \caption{2003 November 18,
M3.9 flare~--~Panels \textbf{a)~--~d)}: Temporal evolution of the flare ribbons in
TRACE~1600~{\AA}. \textbf{a)}: impulsive phase image, \textbf{b)}: time of RHESSI
maximum, \textbf{c)}: time of GOES maximum, \textbf{d)}: decay phase image.
\textbf{e)}: calculated total flare area on decay phase image (black with white
contours: negative polarity, gray with black contours: positive polarity),
\textbf{f)}: total flare area on MDI magnetogram. Contours are the same as in e).
MDI data range scaled to $\pm 500$~G out of [$-1700, +1220$]~G.~--~FOV: $320\arcsec
\times 250\arcsec$.} \label{fig:02}
\end{figure}

\begin{figure}[h]
\resizebox{\hsize}{!}{\includegraphics{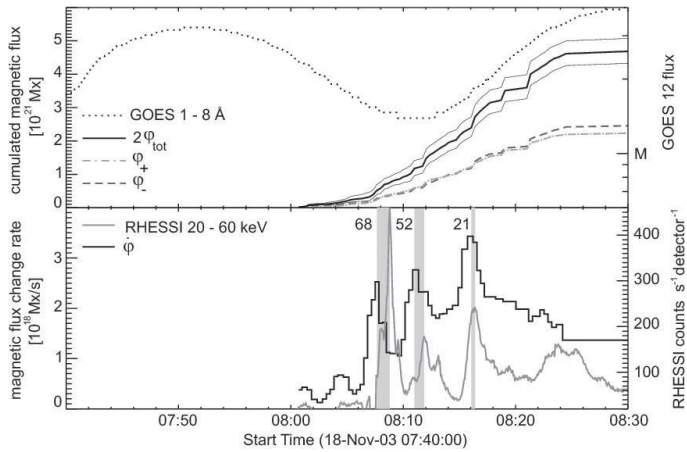}} \caption{2003 November 18, M3.9 flare~--~\textit{Top:} Same as in Fig.~\ref{fig:01a}. The GOES-profile before 08:10~UT shows the occurrence of a previous flare (GOES class M3.2). \textit{Bottom:} RHESSI HXR 20~--~60~keV time profile and magnetic flux change rate
($\dot{\varphi}$). Light-gray vertical bars mark time delays of RHESSI peaks compared to the associated magnetic flux change-rate peaks. Delay in seconds is given on the left of each bar.} \label{fig:02a}
\end{figure}

Panels (a)~--~(d) of Fig.~\ref{fig:02} show the flare evolution in
TRACE~~1600~{\AA}. In Fig.~\ref{fig:02}~(panel~e), the total flare area is presented, and in
Fig.~\ref{fig:02}~(panel~f) the contours of this area are superimposed on the MDI magnetogram.
The northern ribbon, which sweeps across the negative polarity, stops when it
reaches stronger fields. Thus, the flare area (white contours) is forced into a
complex shape by the magnetic field topology. The southern ribbon is not affected by strong magnetic fields, so its shape remains relatively simple, and the positive-polarity flare area (black contours) appears as an elongated structure.

The cumulated positive and negative reconnection-flux profiles look similar (cf.~top panel
of Fig.~\ref{fig:02a}), although, slightly more negative than positive flux is
detected near the end of the analyzed time interval. The ratio of positive to negative flux is 0.91 at about 08:30~UT (cf.~Table~\ref{tab:3}). The total cumulated flux adds up to $2.3 \times 10^{21}$~Mx. We note that the cumulated fluxes of the entire event may be higher than listed in Table~\ref{tab:3}, since it was not possible to analyze the event up to the end of
the impulsive phase due to gaps in the TRACE data after $\sim$~08:24~UT.

In the bottom panel of Fig.~\ref{fig:02a}, the RHESSI HXR 20~--~60~keV time
profile is presented along with the derived magnetic reconnection rate. During the event, RHESSI
observed four major peaks. Three of them are also evident in the
$\dot{\varphi}$-profile, the time interval of a fourth peak being unable to be
analyzed because of the aforementioned gaps in the TRACE data. The co-temporal occurrence of
the first three RHESSI and $\dot{\varphi}$-peaks is, however, obvious, the RHESSI
peaks being slightly delayed, as indicated by the gray vertical bars, highlighting
the time lag between the corresponding peak values. The numbers on the left of
each bar indicate the time delay of the RHESSI peaks in seconds. We note that in
\citet{mikl07}, we did not divide the magnetic field by the cosine of the central
meridional distance of the flare, and therefore obtain in the present paper slightly higher values in the
$\dot{\varphi}$-profile.

\subsection{2005 January 15, X2.6 flare}\label{subsec:15-Jan-2005}

\begin{figure}[h]
\resizebox{\hsize}{!}{\includegraphics{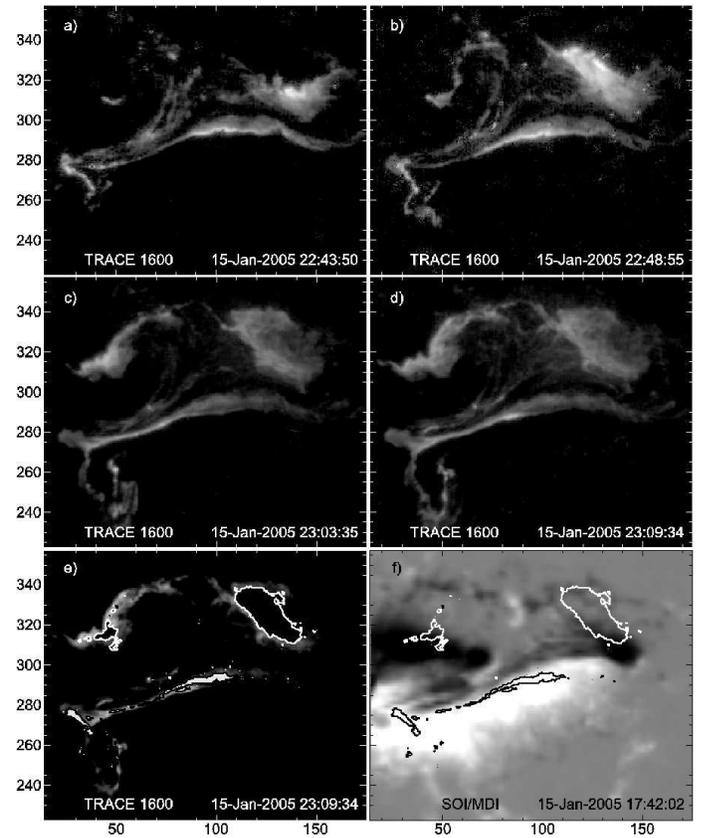}} \caption{2005 January 15,
X2.6 flare~--~Panels \textbf{a)~--~d)}: Temporal evolution of the flare ribbons in
TRACE~1600~{\AA}. \textbf{a)}: impulsive phase image, \textbf{b)}: time of RHESSI
maximum, \textbf{c)}: time of GOES maximum, \textbf{d)}: decay phase image.
\textbf{e)}: calculated total flare area on decay phase image (black with white
contours: negative polarity, white with black contours: positive polarity),
\textbf{f)}: total flare area on MDI magnetogram. Contours are the same as in e).
MDI data range scaled to $\pm 1000$~G out of [$-1470, +1790$]~G.~--~FOV: $160\arcsec
\times 134 \arcsec$.}\label{fig:03}
\end{figure}

Panels (a)~--~(d) of Fig.~\ref{fig:03} show the flare evolution in
TRACE~~1600~{\AA}. Figure~\ref{fig:03}~(panel~e) displays the total flare area, and,
in Fig.~\ref{fig:03}~(panel~f) the area contours are plotted superimposed on the magnetogram of the flaring
region. The negative-polarity area (white contours) is large compared to the
positive one (black contours). This difference in flare area is caused by the negative-polarity ribbon sweeping, at least partly, areas of weaker magnetic field, whereas the positive-polarity ribbon occupies exclusively strong
field regions, some of them exceeding 1500~G. This produces comparable fluxes, although the size of the flare areas differ.

\begin{figure}[h]
\resizebox{\hsize}{!}{\includegraphics{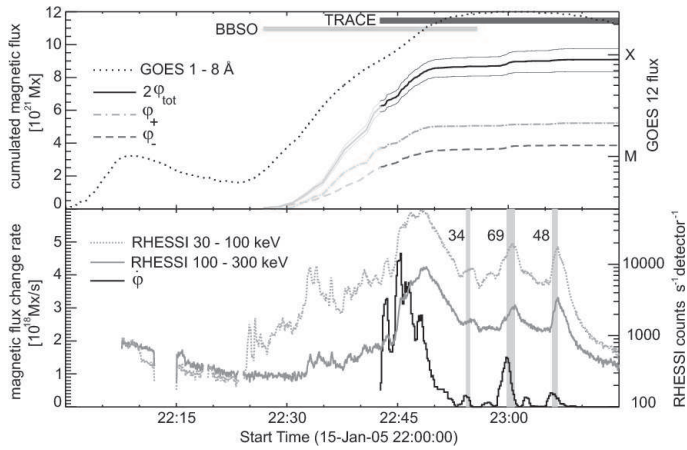}} \caption{2005 January 15, X2.6 flare~--~\textit{Top:} Gray and black horizontal bars mark the
instrument observation times for BBSO H$\alpha$ and TRACE~1600~{\AA}. Profiles
derived from TRACE observations begin at $\sim$~22:42~UT and were attached to the
light-gray profiles obtained from BBSO observations. \textit{Bottom:} RHESSI HXR
30~--~100~keV and 100~--~300~keV time profiles and magnetic flux change rate
($\dot{\varphi}$). \emph{Light-gray vertical bars} mark time delays of RHESSI
100~--~300~keV peaks compared to the associated magnetic flux change-rate peaks.
Delay in seconds is given on the left of each bar.} \label{fig:03a}
\end{figure}

The horizontal bars in the top panel of Fig.~\ref{fig:03a} indicate the instrument
observation times of BBSO and TRACE. Since TRACE observations did not start before
$\sim$~22:42~UT, we used BBSO~H$\alpha$ images, which covered the entire impulsive
phase, to estimate the cumulated reconnection fluxes for the time interval in which
TRACE data was missing. We calculated the fluxes separately for H$\alpha$ and
TRACE~1600~{\AA}, and then we attached the flux profiles obtained from TRACE images
to those derived from BBSO images up to the start of TRACE observations.
Since the H$\alpha$ images were saturated, the fluxes derived from BBSO may be
overestimated, although, they still provide a reference level for the beginning of
the TRACE profiles, and thus allow the estimation of the overall cumulated flux in
this event. The curve progression of the positive and negative fluxes is similar,
yet, we detect more positive than negative flux during this event. At the end of
the analyzed time interval, the ratio of cumulated positive to negative flux is
1.35, and the total flux $\varphi_{\rm{tot}}$ adds up to $\sim 4.5 \times 10^{21}$~Mx
(cf. Table~\ref{tab:3}).

The magnetic flux change-rate derived from TRACE images along with two RHESSI
light curves are shown in the bottom panel of Fig.~\ref{fig:03a}. The RHESSI peak
between $\sim$~22:43~UT and 22:53~UT is unusually broad, and in the 30~--~100~keV and 100~--~300~keV profiles the highest RHESSI peak appears later than the maximum in $\dot{\varphi}$. Until $\sim$~22:55~UT,
the magnetic flux change rate exhibits four spikes embedded in a broad peak, while
RHESSI exhibits the aforementioned broad main-peak. Thus, we did not assign the main
RHESSI peak to any of the $\dot{\varphi}$-peaks. The correspondence between the remaining RHESSI and
$\dot{\varphi}$-peaks is more obvious. After $\sim$~22:55~UT, the
$\dot{\varphi}$-profile and the progression of the nonthermal emission look similar, although the relative height of the RHESSI peaks is not always reproduced by the $\dot{\varphi}$-profile, e.g., the last RHESSI peak at about 23:07~UT is
high in comparison to its counterpart in the magnetic flux change rate. The co-temporal occurrence of observed and derived peaks is also obvious, however, in this event, the RHESSI peaks being slightly delayed. The vertical
bars in the bottom panel of Fig.~\ref{fig:03a} indicate the time lag between the
peaks in the RHESSI 100~--~300~keV and $\dot{\varphi}$-profiles, and the numbers
on the left of each bar indicate the time delay in seconds. We also checked
the delay times of RHESSI peaks in the 30~--~100~keV profile and found delays of
53, 54, and 48~s. Both sets of delay times are comparable to those found in the 2003
November 18 event, presented in Sect.~\ref{subsec:18-Nov-2003}.

\subsection{2006 July 06, M2.5 flare}\label{subsec:06-Jul-2006}

\begin{figure}[h]
\resizebox{\hsize}{!}{\includegraphics{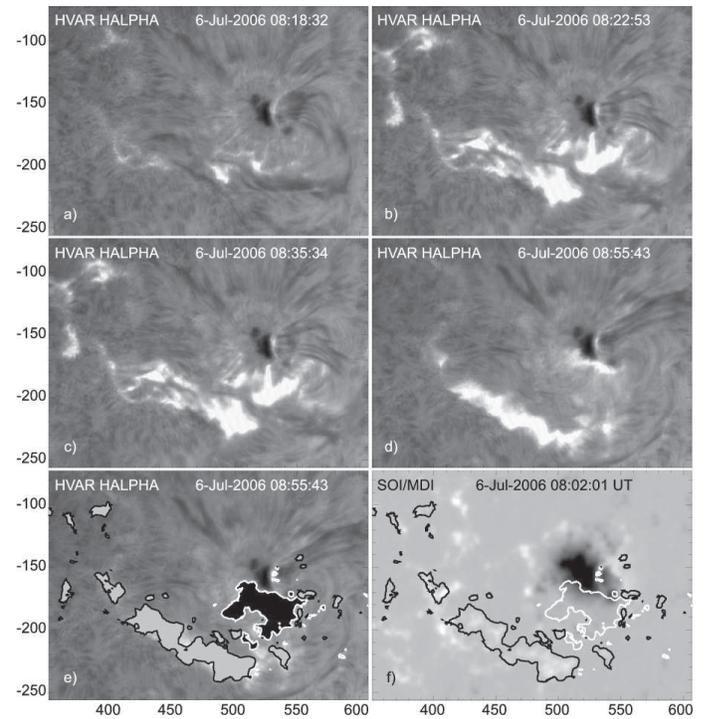}} \caption{2006 July 06,
M2.5 flare~--~Panels \textbf{a)~--~d)}: Temporal evolution of the flare ribbons in
H$\alpha$. \textbf{a)}: impulsive phase image, \textbf{b)}: time of RHESSI
maximum, \textbf{c)}: time of GOES maximum, \textbf{d)}: decay phase image.
\textbf{e)}: calculated total flare area on decay phase image (black with white
contours: negative polarity, white with black contours: positive polarity),
\textbf{f)}: total flare area on MDI magnetogram. Contours are the same as in e).
MDI data range scaled to [$-1000, +400$]~G out of [$-1390, +450$]~G.~--~FOV: $255\arcsec
\times 183 \arcsec$.} \label{fig:04}
\end{figure}

Panels (a)~--~(d) of Fig.~\ref{fig:04} show the flare evolution observed in Hvar
H$\alpha$, which was associated with an erupting filament. Figure~\ref{fig:04}~(panel~e) displays the
total flare area, and in Fig.~\ref{fig:04}~(panel~f) the area contours are
superimposed on the MDI magnetogram. The negative-polarity area is small compared
to the positive one, because the northern ribbon enters the penumbra of the nearby
sunspot, where the magnetic fields are strong. However, it stops at the
border between penumbra and umbra.

\begin{figure}[h]
\resizebox{\hsize}{!}{\includegraphics{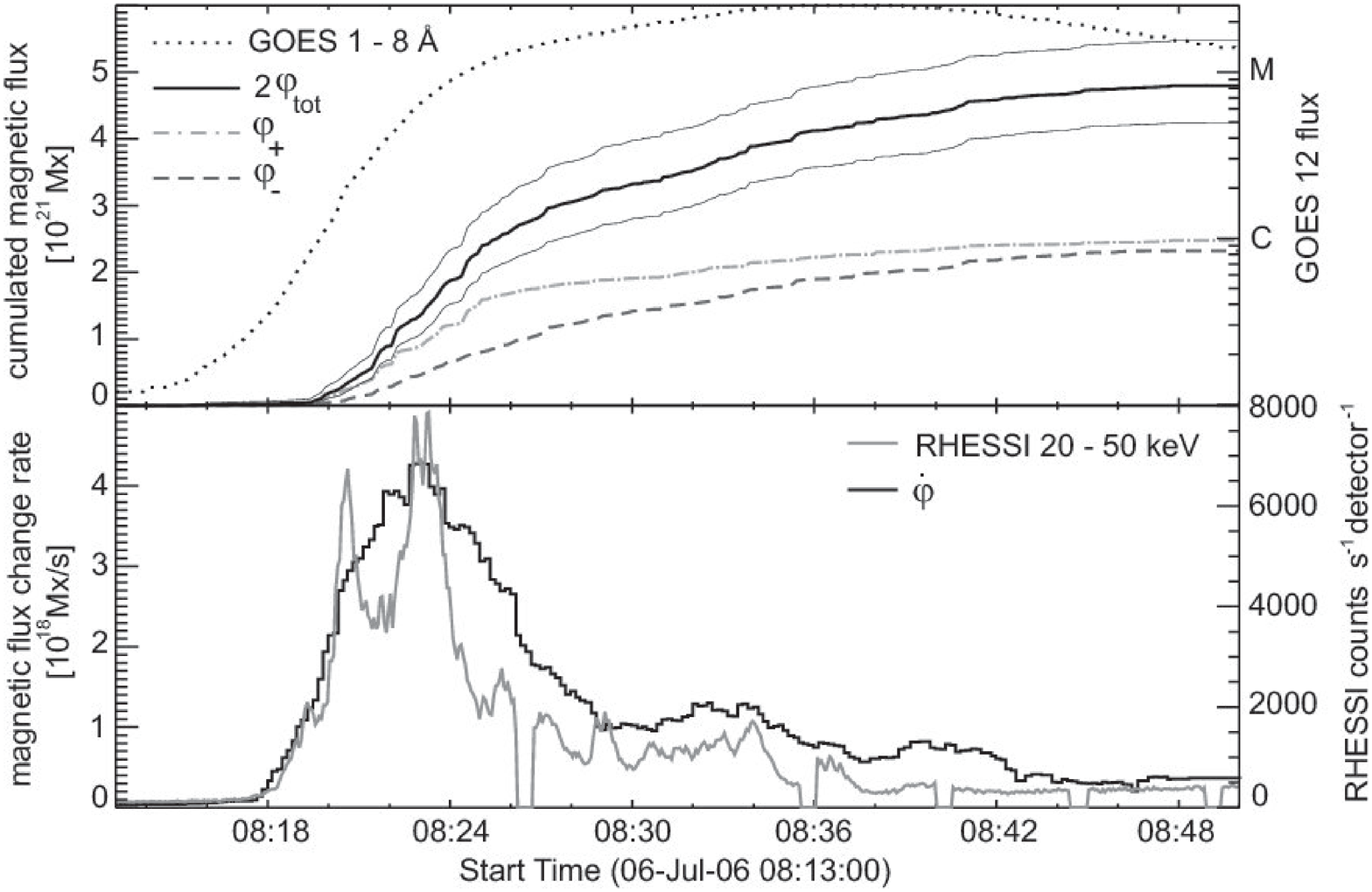}} \caption{2006 July 06,
M2.5 flare~--~\textit{Top:}  Same as in Fig.~\ref{fig:01a}. \textit{Bottom:}
Magnetic flux change rate ($\dot{\varphi}$) and RHESSI HXR 20~--~50~keV time
profile. Counts at times of RHESSI shutter movements are set to 0.} \label{fig:04a}
\end{figure}

In this event, more positive than negative reconnection flux is detected over the
entire analyzed time range (cf.~top panel of Fig.~\ref{fig:04a}). During the first part of the impulsive phase in particular, the positive cumulated flux profile exhibits a steep rise, while the negative flux increases more gradually, its
final flattening occurring later than that of the cumulated positive flux. Thus,
cumulated positive and negative fluxes become comparable in strength at the end
of the analyzed time interval, where the flux ratio is 1.06, and the total
cumulated flux adds up to $2.6 \times 10^{21}$~Mx (cf.~Table~\ref{tab:3}). We note
that the H$\alpha$ images were saturated during the impulsive phase of the event, so the calculated flux values may be
overestimated.

In the bottom panel of Fig.~\ref{fig:04a}, the RHESSI 20~--~50~keV time profile
along with the derived reconnection rate are plotted. In this event, the
nonthermal emission exhibits two distinct peaks between 08:18~UT and 08:25~UT. Afterwards, several small spikes are visible between $\sim$~08:28~UT and 08:37~UT. The two main RHESSI peaks are not resolved in the
magnetic-flux change rate. In the $\dot{\varphi}$-profile, only one broad peak
is evident, which spans the time range of the two main RHESSI peaks. The maximum of
the broad $\dot{\varphi}$-peak corresponds with the second RHESSI peak, and the RHESSI spikes occurring after 08:28~UT is evident in the magnetic-flux change rate as a broad bump. However, there is a second lower bump in the
$\dot{\varphi}$-profile after 08:37~UT, where no corresponding increase in the
nonthermal emission is visible.

\subsection{2007 May 19, B9.5 flare}\label{subsec:19-May-2007}

\begin{figure}[h]
\resizebox{\hsize}{!}{\includegraphics{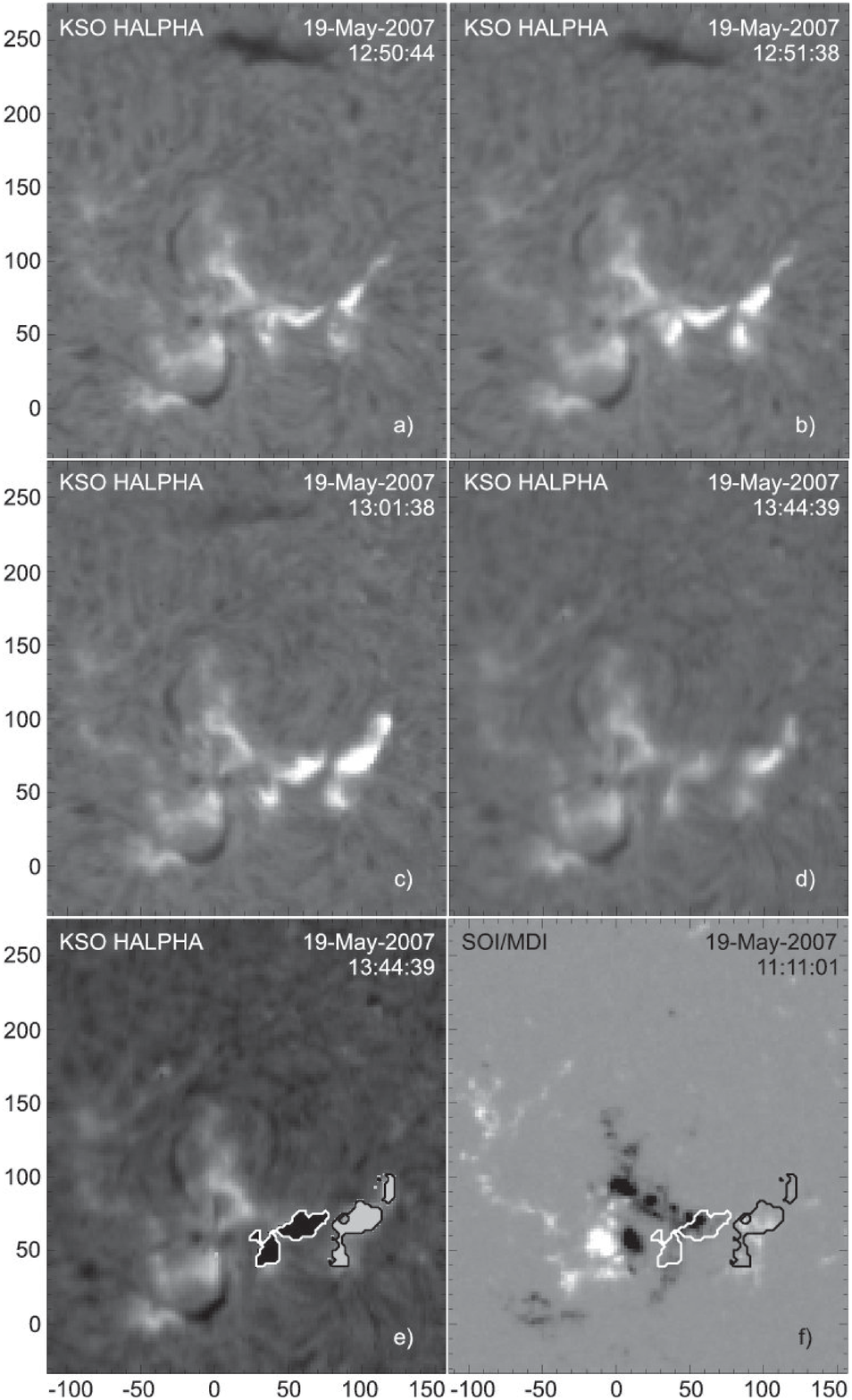}} \caption{2007 May 19,
B9.5 flare~--~Panels \textbf{a)~--~d)}: Temporal evolution of the flare ribbons in
H$\alpha$. \textbf{a)}: impulsive phase image, \textbf{b)}: time of RHESSI
maximum, \textbf{c)}: time of GOES maximum, \textbf{d)}: decay phase image.
\textbf{e)}: calculated total flare area on decay phase image (black with white
contours: negative polarity, gray with black contours: positive polarity),
\textbf{f)}: total flare area on MDI magnetogram. Contours are the same as in e). MDI data range scaled to $\pm 1000$~G
out of [$-1900, +1910$]~G.~--~FOV: $268\arcsec \times 306 \arcsec$.} \label{fig:05}
\end{figure}

Panels (a)~--~(d) of Fig.~\ref{fig:05} show the flare evolution observed in KSO H$\alpha$. There are two stable filaments east of the flare site. Two further filaments in this active region were erupting \citep[cf.][]{veronig08}. The first one, whose southern footpoint was anchored near the flare site, went off towards west and has already disappeared from the H$\alpha$ filter at 12:50~UT. A part of the second erupting filament is still visible as a dark, elongated structure in panels (a) and (b) at the top of the images, although it is already in a state of eruption. In panel (c), it has almost left the H$\alpha$ filter, and in Fig.~\ref{fig:05}~(panel~d) it has completely vanished. Figure~\ref{fig:05}~(panel~e) shows the total flare area in this event, and in Fig.~\ref{fig:05} (panel~f) the area contours are plotted on the MDI magnetogram. The size of
the positive and negative-polarity areas is comparable. Both areas are elongated structures.

\begin{figure}[h]
\resizebox{\hsize}{!}{\includegraphics{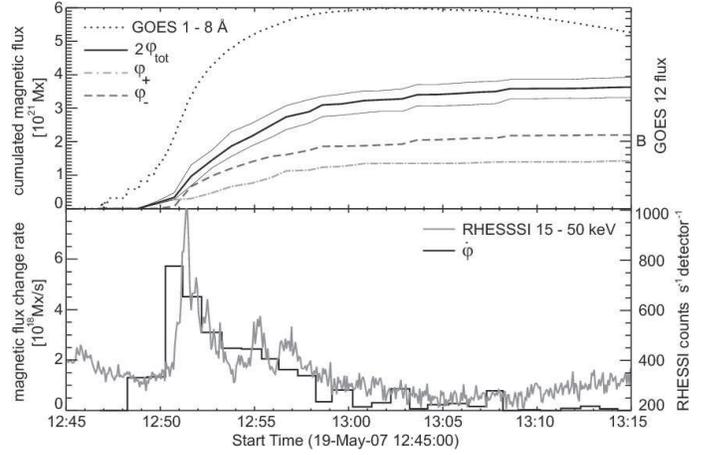}} \caption{2007 May 19,
B9.5 flare~--~\textit{Top:} Same as in Fig.~\ref{fig:01a}. \textit{Bottom:} RHESSI
HXR 15~--~50~keV time profile and magnetic flux change rate ($\dot{\varphi}$).}
\label{fig:05a}
\end{figure}

In this comparatively weak event, more negative than positive reconnection flux is
detected (cf. top panel of Fig.~\ref{fig:05a}). At the end of the analyzed time
interval, the ratio of cumulated positive to negative flux is 0.64. The total flux
adds up to $1.8 \times 10^{21}$~Mx (cf.~Table~\ref{tab:3}). We note that the H$\alpha$
images were saturated during the impulsive phase of the event, so the calculated flux values may be
overestimated.

\begin{figure*}
\resizebox{\hsize}{!}{\includegraphics{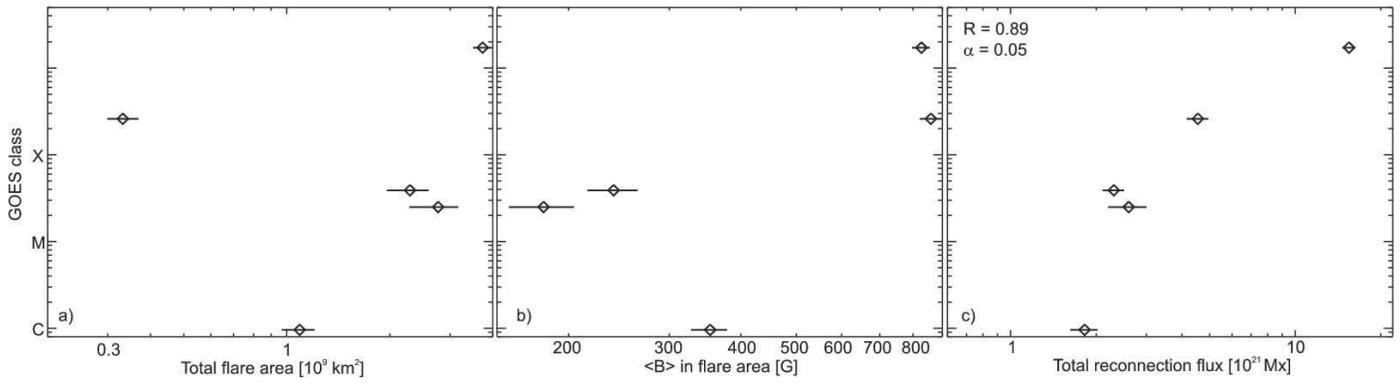}} \caption{GOES peak flux vs.:~\textbf{a)} Total flare area $A$; \textbf{b)} Mean magnetic field strength $\langle B\,\rangle$ inside $A$; \textbf{c)} Total reconnection flux. In the upper left corner of c) the Pearson correlation coefficient $R$, along with $\alpha$, the probability of error, is shown. Error bars are obtained from the lowest and highest of the appropriate intensity thresholds (see
Sect.~\ref{sec:analysis} for more detail). Error bars in (b) also account for the MDI noise level of 20~G.}
\label{fig:06}
\end{figure*}

The bottom panel of Fig.~\ref{fig:05a} displays the derived magnetic-flux change
rate, as well as the RHESSI 15~--~50~keV time profile. Although the lower end of
this energy band could in principle be thermal emission, the shape of the RHESSI
peaks and the RHESSI spectra indicate that the emission is predominantly nonthermal. The first and highest RHESSI peak is also evident in the $\dot{\varphi}$-profile, whereas the second and third peak are unresolved.

\begin{table}
\begin{center}
\caption{Cumulated reconnection fluxes and ratio of positive to negative flux at
the end of the analyzed time interval. Fluxes are given in units of $10^{21}$~Mx.
Fluxes in the flare on 2003 November 18, may be higher, since it was not possible
to analyze the event up to the end of the impulsive phase due to gaps in the TRACE
data after $\sim$~08:24~UT. Error estimates in $\varphi_{\rm{tot}}$ are obtained from
the lowest and highest of the appropriate intensity thresholds (see
Sect.~\ref{sec:analysis} for more detail).} \label{tab:3}
\begin{tabular}{c c c c c c c}
  \hline \hline
    & GOES &  &  &   &   \\
  \raisebox{1.2ex}[-1.2ex]{Date} & class & \raisebox{1.2ex}[-1.2ex]{$\varphi_+$}
    & \raisebox{1.2ex}[-1.2ex]{$\varphi_-$}& \raisebox{1.2ex}[-1.2ex]{$\varphi_{\rm{tot}}$}
    &  \raisebox{2ex}[-2ex]{$\frac{\varphi_+}{|\varphi_-|}$}   \\
  \hline
    2003 Oct 28 & X17.2& 15.6 & -15.3  & 15.5 $\pm 0.8$ & 1.02\\
    2003 Nov 18 & M3.9 & $>2.2$ & $>-2.4$  & $>2.3$ $\pm 0.2$& 0.91\\
    2005 Jan 15 & X2.6 & 5.2 & -3.9  & 4.5 $\pm 0.4$& 1.35\\
    2006 Jul 06 & M2.5 & 2.7 & -2.5  & 2.6 $\pm 0.4$& 1.06\\
    2007 May 19 & B9.5 & 1.4 & -2.2  & 1.8 $\pm 0.2$& 0.64\\
  \hline
\end{tabular}
\end{center}
\end{table}

\subsection{Relations between reconnection flux, GOES class, and CME
velocity}\label{subsec:relations}

In Fig.~\ref{fig:06}, for the five flares studied we plot the measured GOES peak flux versus.:~a)~the total flare
area $A$, i.e., the sum of the newly brightened areas in all images; b)~the mean
magnetic field strength $\langle B\,\rangle$ inside $A$; and c)~ the total magnetic
reconnection flux $\varphi_{\rm{tot}}$. We calculated the three quantities for each of the appropriate intensity thresholds (see Sect.~\ref{sec:analysis}) and plot the average. Error bars were obtained from $A$, $\langle B\,\rangle$, and $\varphi_{\rm{tot}}$ for the lowest and highest thresholds. Error bars in Fig.~\ref{fig:06}~(panel~b) also include the effects of the MDI noise level of 20~G. After calculating in \emph{log-log}-space the
correlation between the GOES peak of each event and $A$, $\langle B\,\rangle$, and
$\varphi_{\rm{tot}}$, respectively, the GOES peak was found to neither correlate
with $A$ nor $\langle B\,\rangle$, although, it is correlated significantly with
$\varphi_{\rm{tot}}$ ($R = 0.89$, confidence level 95\%, cf.~Fig.~\ref{fig:06}~(panel~c)). Since our sample size is small, we note that  we cannot exclude the possibility that $A$ and $\langle B\,\rangle$ are also related to the GOES peak flux (see Sect.~\ref{sec:discussion2} for a detailed discussion on this issue).

\begin{figure}[h]
\resizebox{8cm}{!}{\includegraphics{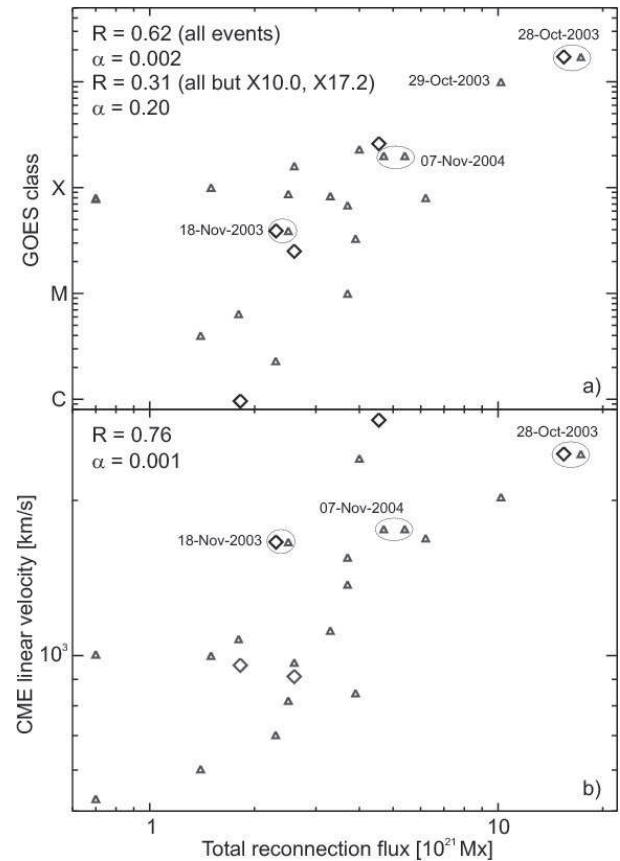}} \caption{\textbf{a)} GOES peak flux vs.~$\varphi_{\rm{tot}}$; \textbf{b)} Linear CME velocity, taken from the SOHO LASCO CME catalog, vs.~$\varphi_{\rm{tot}}$. $\diamond$ mark events from this study (cf. Fig.~\ref{fig:06}c); $\triangle$ indicate events taken from \citet{qiu05}, \citet{qiu07}, and \citet{longcope07}. Same events analyzed in different studies are encircled. Upper left corner of each panel shows Pearson correlation coefficient $R$ along with $\alpha$, the probability of error.}
\label{fig:06a}
\end{figure}

In Fig.~\ref{fig:06a}~(panel~a), we combine the reconnection fluxes from the five events
analyzed in this paper with results from other events, derived by \citet{qiu05},
\citet{qiu07}, and \citet{longcope07}. We note that these authors applied potential-field extrapolation from the photospheric MDI magnetograms to a height of 2~Mm, which we did not. Nevertheless, our results are consistent with the enlarged data set, and we derive comparable reconnection fluxes for the events on 2003 October 28 and 2003
November 18 (cf. encircled events in Fig.~\ref{fig:06a}). The combined data
set contains two events exceeding GOES class X10 (2003 October 28 and 29). For the entire data set,
the correlation between GOES peak and $\varphi_{\rm{tot}}$ is significant ($R=0.62$, with a confidence level greater than 99\%). However, if the two most energetic events are excluded, the correlation ($R=0.31$) is
significant only with a confidence level of 80\%. This indicates that the rare $\geq
\rm{X}10$-events strongly contribute to the correlation.

We also investigated the relation between the total reconnection flux in the flare and the kinematics of the associated CME. In Fig.~\ref{fig:06a}~(panel~b), we plot the linear CME velocity, taken from the SOHO LASCO
CME catalog \citep{yashiro04}, i.e., the linear speed obtained by fitting a straight line to the
height-time measurements, against $\varphi_{\rm{tot}}$ for all events from
Fig.~\ref{fig:06a}~(panel~a). Using only the five events analyzed in this paper (marked as
$\diamond$ in Fig.~\ref{fig:06a}~(panel~b)), the correlation ($R=0.71$) is significant only with an 80\%-confidence level, although, taking the enlarged data set, we find a correlation that is significant with a confidence level greater than 99\% (cf.~Fig.~\ref{fig:06a}~(panel~b)), the correlation coefficient ($R=0.76$) being lower, however, than that reported by \citet{qiu05} for the subset of 13~events analyzed by these authors. Nevertheless, our result is in line with that of \citet{qiu05},
indicating that flares of higher reconnection flux are associated with faster CMEs.

\section{Discussion}\label{sec:discussion}


In the following, we list potential errors in the determination of the 
magnetic reconnection flux that may arise from the applied methods, as well as the
technical equipment and data acquisition procedure. Afterwards, we discuss the
physical implications of our results.

\subsection{Potential errors}

Potentially significant measurement errors, which can result from transient bright
non-ribbon features, such as cosmic rays, were accounted for by eliminating short-lived features from the images (see Sect.~\ref{sec:dobs} for more detail). Disregarding transient features can result in an overestimation
by 100\% of the total flare area and 20\% of the total reconnection flux, if an
image time series is strongly corrupted by particle hits. Bright coronal loops may also overestimate the flare area, and thus, the
reconnected flux, when counted among flare pixels. Therefore, we excluded these
regions from the analysis by setting the corresponding pixels during the time
range, when these brightenings occurred, to zero.

A crucial factor is the threshold value, which is used to discern flare pixels and
non-flare pixels. \citet{longcope07} reported a change of 25\% in the
measured flux, when experimenting with cutoff values. We also tested several
cutoff values and found a change of 5~--~15\% in the measured reconnection flux.

A further effect, which may cause an overestimation of the flare area, and thus,
the reconnection flux, is caused by a limit to the amount of
charge that each pixel of a CCD chip can store. If there is too much charge at a
particular location on the CCD chip, i.e., if a pixel is saturated, it will
overflow to its neighboring pixels, preferentially the pixels above and below the
saturated pixel. The signal in those pixels is questionable, since they will usually
contain spilled charge. In three of the events (2005 January 15, X2.6, 2006 July 06, M2.5, and 2007 May 19, B9.5), H$\alpha$ images were saturated during the impulsive phase, and the area and reconnection flux may thus be overestimated.

All sources of uncertainty mentioned so far are related to the flare area. The
second quantity required to determine magnetic reconnection rates and
fluxes, namely, the magnetic field, is measured routinely only in the
photosphere. Thus, photospheric line-of-sight magnetograms are commonly used for
this purpose. However, since the magnetic field strength decreases from the
photosphere to the chromosphere/transition region, where the flare area is
measured, reconnection fluxes are overestimated, when the magnetic field is derived
from photospheric magnetograms, as was done for the present paper. \citet{qiu07}
reported an overestimation of about 20\% in the reconnection flux derived from
photospheric magnetic fields compared to the flux derived using potential field
extrapolation to a height of 2~Mm.

As for the magnetic field, a further potential source of error must be accounted for.
\citet{berger03} reported that MDI measurements underestimate fields stronger than
1700~G, and \citet{qiu07} estimated an upper limit to possibly underestimated
reconnection flux due to MDI saturation effects, assuming that saturation occurs at field strengths of 1700, 1600, and 1500~G. In two of the events analyzed in this paper (2003 October 28, X17.2 and 2005 January 15, X2.6), small parts of the flare
ribbons entered regions with field strengths stronger than 1500~G. According to
\citet{qiu07}, this may result in an underestimate of a few percent, at the very
maximum, of the total reconnection flux. To account for this effect, we followed the cross-calibration study of \citet{berger03} and multiplied magnetic field values by a correction factor of 1.56.

The misalignment between UV/H$\alpha$ images and MDI magnetograms is another
source of uncertainty, which can be estimated by artificially offsetting the two
sets of images. We did not experiment with intentionally misaligned images, but
\citet{longcope07} and \citet{qiu05} reported that artificial misalignment by up
to 4\arcsec~can account for up to 10~--~20\% of change in the measured flux. However,
the co-alignment accuracy that we achieved was higher than 2\arcsec.

Bearing in mind all the sources of uncertainty discussed above, it is difficult to provide an overall error estimate for the total reconnection flux in each of the five events analyzed in this paper. Effects of over- and
underestimation may cancel each other to a certain degree. To be on the safe
side, an overestimate of 30~--~40\% should be taken into account, especially for events in which the fluxes were derived from saturated H$\alpha$ images.

\subsection{Consideration of the results in view of the standard flare/CME model and numerical MHD models}\label{sec:discussion2}

We derived the magnetic-flux change rate from chromospheric/photospheric
observations of five flare events (GOES classes B, M, and X) and compared it to the
observed nonthermal flare emission. In addition, we calculated the cumulated
positive and negative magnetic flux participating in the reconnection
process, as well as the total reconnection flux $\varphi_{\rm{tot}}$. We also
determined the correlation between the $\varphi_{\rm{tot}}$, GOES peak, and linear velocity
of the flare-associated CMEs.

We found good temporal correlations between the derived magnetic-flux change-rate and
observed nonthermal emission in all events, i.e., hard X-ray peaks were clearly
reflected in the magnetic flux change-rate profiles, as expected from the
standard model, although the relative height of the peaks was not always
reproduced in the $\dot{\varphi}$-profiles. In two events, for which non-saturated
TRACE image time series were available with particularly high cadence ($\leq 23$~s), the
HXR peaks appeared delayed by roughly 1~min. Considering the amount of this delay,
in \citet{mikl07}, we speculated that it might be related to the travel time of a
reconnected field line from the diffusion region to the lower edge of the current
sheet. However, since we found a delay of HXR peaks in only two events so far, no
conclusions can be drawn at present. It is, however, interesting to note that \citet{warmuth09} calculated a travel time of $92$~s in the 2003 November 18 flare, when comparing their model of shock drift acceleration at the reconnection outflow termination shock with observations.

The derived total reconnection fluxes ranged between $1.8 \times 10^{21}$~Mx for
the weakest event and $15.5\times 10^{21}$~Mx for the most energetic one
(cf.~Table~\ref{tab:3}). According to the standard model, equal amounts of positive
and negative magnetic flux should participate in the reconnection process. The
cumulated positive and negative fluxes that we found were roughly balanced, with
flux ratios ranging between 0.64 and 1.35. According to \citet{qiu05}, flux ratios of
between 0.5 and 2 can be regarded as a good flux balance, given the numerous
uncertainties involved in the measurements.

Beside determining reconnection rates and fluxes, we also calculated the
correlation between GOES peak flux and: (1) total flare area; (2) mean magnetic
field strength inside this area; and (3) total reconnection flux. Although neither the
flare area nor the magnetic field were correlated with the GOES class, the total reconnection flux was. However, we note that our sample size is small. Therefore, we cannot exclude the possibility that the other two parameters are also related to the GOES peak flux, and statistical studies have indeed found such correlations. \citet{nagashima06} analyzed 77 C, M, and X flares and detected a threshold value in the magnetic field that increases with the GOES peak flux. For example, X-class flares occurred only when the mean magnetic field in the active region was stronger than $\sim 100$~G. \citet{su07a} analyzed a sample of 31 flares and found that for events of stronger average magnetic field strength, the GOES peak flux tended to be higher. In a subsample of 18 out of the 31 events, they also detected a positive correlation between the GOES class and both the flare area and the magnetic flux, the magnetic flux showing a much stronger correlation than the area or the magnetic field. \citet{su07a} attributed this to the magnetic flux being the product of the other two parameters. Although the applied methods for determining the flare area differ from our approach~--~\citet{nagashima06} used intensity thresholds in soft X-ray images, and \citet{su07a} took a contour level in photospheric magnetograms of the flaring region as a basis for the calculation of $A$~--~our findings are in line with the explanation given by \citet{su07a}, because they also indicate that in this context the combination of area and magnetic field inside this area is important. It is conceivable that we did not find the weaker correlations between GOES class and $A$ and $\langle B\,\rangle$ due to the small size of our sample, but we did find the stronger correlation between GOES peak flux and $\varphi_{\rm{tot}}$.

In addition, we incorporated our flux results into a larger set of results, taken from \citet{qiu05}, \citet{qiu07}, and \citet{longcope07}, and again found a
significant correlation between GOES peak flux and total reconnection flux ($R=0.6$, confidence level greater than 99\%).
However, the correlation became smaller ($R=0.3$), and was significant only with an 80\%-confidence level, when the most energetic events (GOES classes $\geq \rm{X}10$) were excluded. This
suggests that the rare, most energetic flares strongly contribute to the correlation. Still, the results indicate that the amount of magnetic flux participating in the reconnection process is larger in more energetic events than in weaker ones, as theoretically expected. The more magnetic flux is reconnected in a flare, the more energy is released into fast particles and can subsequently be deposited into the chromosphere. This energy heats the chromospheric plasma, which then evaporates, fills the flare loops, and causes them to emit soft X-ray radiation, measured by the GOES satellites. The event is assigned to a particular class in the GOES classification scheme according to its maximum soft X-ray emission, which can be understood as a measure of the cumulated energy in the hot thermal plasma \citep[e.g.,][]{veronig02,veronig05}.

We also found that flares with more reconnection flux are associated with
faster CMEs. This correlation, which was also reported by \citet{qiu05}, can be explained by a feedback relationship between the CME kinematics and the reconnection process in the flare \citep{bojan05c,maricic07,temmer08}. According to the standard flare/CME model, 50\% of the magnetic flux reconnected during a dynamical (two-ribbon) flare is transported upward to higher coronal heights and even interplanetary space. The reconnection process in the flare, which is initiated in the wake of the CME, affects the CME by: (1) reducing the net tension of the overlying magnetic field; (2) increasing the magnetic pressure below the erupting flux-rope; and (3) supplying the flux-rope with extra poloidal flux \citep{maricic07}. This enhances and prolongs the acceleration of the flux-rope and therewith additionally drives the CME. Therefore, CMEs that are associated with dynamical flares are generally faster than those linked with eruptive filaments, and CMEs that are associated with more powerful flares, i.e., flares of higher GOES class or fluence (time-integrated SXR flux), are of higher median speed and kinetic energy than weak-flare-associated CMEs \citep{moon02,bojan05c}. In other words, the higher the amount of flux reconnected in a flare, the higher the amount of flux going upward, resulting ultimately in faster CMEs. 

Numerical MHD models dealing with the formation of a flux rope, its sudden eruption and escape from the Sun provide additional evidence of a feedback relationship between an erupting structure and magnetic reconnection. \citet{cheng03} accomplished 2.5-dimensional resistive MHD simulations of the evolution of an initially closed arcade field configuration. By imposing a shear-increasing footpoint motion, magnetic reconnection occurs, creating a new flux rope. The toroidal flux originally contained in the arcade is redistributed into both the slowly rising flux rope and the underlying arcade field. Thus, the magnetic shear in the arcade is reduced after the flux rope formation, but increases again, if additional footpoint motion occurs. As soon as the magnetic shear exceeds a critical value, another reconnection event occurs and a new flux rope is formed. When magnetic reconnection continues, this newborn flux rope rises with increasing velocity, and finally merges with the previously created flux rope. This process of flux rope formation and merging can be repeated as long as magnetic shear is continuously supplied. \citet{cheng03} quantitatively compared the model results with low-corona observations of a filament eruption during a flare/CME event and found good agreement in the evolution of the flux rope height, velocity, and acceleration during the flux rope acceleration phase. Although the model does not address the escape of a flux rope from the Sun into interplanetary space, it shows that reconnection enhances the flux rope acceleration, and thus supports the scenario of a feedback relationship between the erupting structure and the reconnection in the flare.

\citet{lin00} investigated the interaction between an already existing flux rope and magnetic reconnection. In their two-dimensional model, the flux rope experiences a catastrophic loss of equilibrium when the photospheric sources of the coronal magnetic field are brought together quasi-statically and reach a critical distance. Then, the flux rope jumps to a new equilibrium position at a higher altitude, and a vertical current sheet is created below the flux rope. The model shows that without reconnection in the current sheet the magnetic tension force is always strong enough to prevent the flux rope from reaching interplanetary space. However, even a modest reconnection rate (prescribed by the inflow Alfv\'en Mach number $M_{\rm{A}}$) is sufficient to allow its escape. Based on this loss-of-equilibrium model and its extension, which includes gravity in the calculations \citep{lin04,reeves05b}, \citet{reeves06} examined the relevance of reconnection for the flux-rope acceleration by analyzing its impact on the individual forces acting on the flux rope. They found that only the force due to the current sheet was affected considerably by changes in the reconnection rate $M_{\rm{A}}$, while other forces, e.g., gravity, exhibited only minor changes with varying $M_{\rm{A}}$. Slow reconnection rates resulted in longer current sheets, and thus, an increase in the downward force exerted by the current sheet on the flux rope. As a consequence, the total acceleration of the flux rope was decreased in cases of slow reconnection rates. When the reconnection rate was fast, the force due to the current sheet was diminished, because fast reconnection rates dissipate the current sheet. Therefore, the total flux rope acceleration was higher in cases of fast reconnection.

All of the models aforementioned are 2D or 2.5D models, which might be appropriate only in describing the eruption of a very long flux rope, especially when its height and width are much smaller than its length. In 3D models, the flux rope is anchored in the photosphere, and because of its final size, the overlying field can be pushed aside by the erupting structure. Thus, unlike in the 2D-approach, in 3D-model reconnection is not necessarily the main mechanism reducing the tension of the overlying field. However, \citet{bojan08} demonstrated that in a 3D-flux-rope model without reconnection high accelerations are prevented because of the inductive decay of the current in the flux rope as the eruption progresses. On the other hand, including reconnection underneath the rising flux rope provides `fresh' poloidal flux to the rope. This preserves the current in the flux rope, and thus prolongs and enhances its acceleration.

The theoretical work discussed above reveals the importance of magnetic reconnection for the kinematics of the erupting structure. Reconnection in the current sheet occurs during the eruption of the flux-rope/CME, additionally drives the eruption, and may even be essential for a flux-rope to escape from the Sun, i.e., there may be no fast CMEs without magnetic reconnection, which is generally believed to play a major role in solar flares. However, it is known from observations that CMEs are not necessarily associated with flares. Instead, they may occur together with erupting filaments. After the eruption of a quiescent filament a growing system of post-eruption loops is often observed in the EUV range \citep[e.g.,][]{bojan05c}. Morphologically, these loop systems are similar to postflare loops, except for the fact that they are not hot enough to be seen in soft X-rays. Quiescent filament eruptions occur in quiet regions, where the plasma-to-magnetic-pressure ratio $\beta$ is generally larger than in active regions \citep{wu01}. \citet{bojan05} demonstrated that in a $\beta > 0.1$ environment, the reconnection outflow region is not heated much. Therefore, an eruption taking place in such an environment should not be accompanied by a flare.

The loss-of-equilibrium model also indicates that insufficient plasma heating may be responsible for non-flare CMEs. \citet{reeves05} simulated several erupting flux-rope/CME events, using background magnetic fields that ranged from weak to higher fields strengths. They found that weaker/higher background fields resulted in slower/faster CMEs, if all other parameters in the calculations remained the same. They also calculated GOES X-ray light curves for the 1~--~8~{\AA} channel adding two different levels of soft X-ray background flux, namely, the average of the maximum/minimum 1~--~8~{\AA} GOES flux during solar cycle~22. It turned out that with the weakest of the chosen background magnetic fields, adding the low X-ray background level produced an A5 flare, while adding the high X-ray background-flux, the GOES light curve hardly emerged from the background, i.e., in this case, the thermal energy release was not large enough to heat and evaporate sufficient plasma into the loops. Thus, the corresponding GOES light curve progression should probably not be considered to be a flare, although, magnetic reconnection was involved in the eruption of the flux rope/CME.

It is also known from observations that fast CMEs are not always associated with powerful flares \citep[e.g,][]{bojan05c}. Although there are correlations between the kinetic energy/velocity of CMEs and the SXR peak flux in the corresponding flares \citep[e.g.,][]{moon03,burkepile04}, CMEs of comparable speed can by all means be associated with flares of different GOES classes. Two of the flare/CME events analyzed in this paper provide an observational example. The CME-speeds on 2006 July 06 and 2007 May 19 were of the order of $900\,\rm{km\,s^{-1}}$, but the GOES importance of the associated flares was M2.5 and~B9.5, respectively (cf.~Table~\ref{tab:1}). The simulations carried out by \citet{reeves05} also addressed this issue, and the results indicated that the CME speed is not indicative of the amount of associated X-ray flux, i.e., CMEs with very similar trajectories can have quite different flare responses. The authors simulated a high-mass flux rope erupting from a strong background field region as well as a low-mass flux-rope being ejected out of a weaker background field. The resulting velocity profiles for both events were similar. The flux-rope/CME speeds reached values of around $1100\,\rm{km\,s^{-1}}$, although the corresponding GOES peak values, and thus, the importance of the associated simulated flares were different (C5 vs.~M2). \citet{reeves05} attributed the difference in GOES class to the amount of thermal energy released in the current sheet in the model being directly related to the strength of the background magnetic field, while the kinetic energy of the flux rope depends on both the mass of the flux rope and the strength of the background field. Therefore, low mass CMEs from weak-field regions can have a similar velocity profile as massive CMEs from stronger field regions. The difference in the physical properties of both situations is evident only in the simulated X-ray emission of the associated flare.

Taking into account the results provided by numerical MHD simulations of solar eruptions, it becomes apparent that the combination of theoretical work and flare/CME observations is essential for improving our understanding of the underlying physical mechanisms and the relation between the two phenomena.

\begin{acknowledgements}
C.~H.~M.~is grateful to J.~Kiener, H.~Aurass, and A.~Warmuth for providing the
INTEGRAL data. C.~H.~M. also thanks V.~B. Yurchyshyn and the Global High Resolution H$\alpha$ Network, operated by the Big Bear Solar Observatory, New Jersey Institute of Technology, for providing BBSO H$\alpha$ images. C.~H.~M. and A.~M.~V.~gratefully acknowledge the Austrian \emph{Fonds zur F\"orderung der wissenschaftlichen Forschung} (FWF grant P20145-N16) for supporting this project.
\end{acknowledgements}


\end{document}